\documentclass[conference]{IEEEtran}
\ifCLASSINFOpdf
\else
\fi
\usepackage{graphics} 
\usepackage{epsfig} 
\usepackage{mathptmx} 
\usepackage{amsmath} 
\usepackage{amssymb}  
\usepackage{graphicx}
\usepackage{multirow}

\usepackage{tikz}
\usetikzlibrary{calc}
\tikzset{>=latex}
\usetikzlibrary{shapes}

\usepackage{tikz-3dplot}

\usepackage[european,americaninductors,straightlabels]{circuitikz}

\usepackage{pgfplots}
\pgfplotsset{compat=newest}
\pgfplotsset{every axis/.append style={line join=bevel}}
\pgfplotsset{
	every axis/.style={
		width=0.8\textwidth,
		height=0.5\textwidth,
		max space between ticks=50,
	}
}
\pgfplotsset{legend cell align=left}
\pgfplotsset{xmajorgrids}
\pgfplotsset{ymajorgrids}
\pgfplotsset{scale only axis}
\definecolor{matlab1}{rgb}{0,0,1}
\definecolor{matlab2}{rgb}{0,0.5,0}
\definecolor{matlab3}{rgb}{1,0,0}
\definecolor{matlab4}{rgb}{0,0.75,0.75}
\definecolor{matlab5}{rgb}{0.75,0,0.75}
\definecolor{matlab6}{rgb}{0.75,0.75,0}
\definecolor{matlab7}{rgb}{0.25,0.25,0.25}
\pgfplotscreateplotcyclelist{matlab}{
	{matlab1,solid},
	{matlab2,dashed},
	{matlab3,dashdotted},
	{matlab4,dotted},
	{matlab5,densely dashed},
	{matlab6,densely dashdotted},
	{matlab7,densely dotted}
}
\pgfplotsset{cycle list name=matlab}
\pgfplotsset{every axis plot/.append style={line width=1pt}}
\pgfplotsset{/pgf/number format/.cd,1000 sep={\,}}
\pgfplotsset{/pgfplots/legend pos/north/.style={/pgfplots/legend style={at={(0.50,0.97)},anchor=north}}}
\pgfplotsset{/pgfplots/legend pos/south/.style={/pgfplots/legend style={at={(0.50,0.03)},anchor=south}}}
\pgfplotsset{/pgfplots/legend pos/east/.style={/pgfplots/legend style={at={(0.97,0.50)},anchor=east}}}
\pgfplotsset{/pgfplots/legend pos/west/.style={/pgfplots/legend style={at={(0.03,0.50)},anchor=west}}}
\pgfplotsset{/pgfplots/legend pos/outer north/.style={/pgfplots/legend style={at={(0.50,1.03)},anchor=south}}}

\usepackage[binary-units]{siunitx}
\title{
Electric vehicle charging during the day or at night: A perspective on carbon emissions}

\vspace{-0.15in}
\author{\IEEEauthorblockN{Xiao Chen$^{1}$, Chin-Woo Tan$^{1}$, Sila Kiliccote$^{2}$, Ram Rajagopal$^{1}$}
\IEEEauthorblockA{$^{1}$Civil \& Environmental Engineering, Stanford University, Stanford, CA 94305, USA\\
$^{2}$eIQ mobility, Oakland, CA 94607\\
 \texttt{markcx@stanford.edu, tancw@stanford.edu, sila@eiqmobility.com, ramr@stanford.edu}}
}

\begin{document}
\vspace{-0.15in}
\maketitle
\vspace{-0.17in}
%
\begin{abstract}
We propose an emission-oriented charging scheme to evaluate the emissions of electric vehicle (EV) charging from the electricity sector at the region of Electric Reliability Council of Texas (ERCOT). We investigate both day- and night-charging scenarios combined with realistic system load demand under the emission-oriented vs direct charging schemes. Our emission-oriented charging scheme reduces carbon emissions in the day by 13.8\% on average. We also find that emission-oriented charging results in a significant CO$_2$ reduction in 30\% of the days in a year compared with direct charging. Apart from offering a flat rebate for EV owners, our analysis reveals that certain policy incentives (e.g. pricing) regarding EV charging should be taken into account in order to reflect the benefits of emissions reduction that haven't been incorporated in the current market of electricity transactions.
\end{abstract}

%

%
\vspace{-0.11in}
\section{Introduction}
\vspace{-0.01in}
%
Green house gas (GHG) emissions from transportation sector account for 27\% of U.S. annual GHG emissions\cite{usepa2017inventory}. Mitigating these emissions has focused on promoting electric vehicles (EVs) as a solution because EVs have lower tailpipe emissions than gasoline-powered vehicles. Yet EV batteries need to be recharged.
There would be an increase in electricity generation at power plants that provide the marginal power to satisfy the EV charging demand. An accelerated adoption of EVs will affect the emissions from the electricity sector, given the fact that power generation also accounts for over 40\% of GHG emissions\cite{weis2015emissions, eia2009emissions}. Therefore, the quantification of the actual changes in emissions in the electricity sector requires a more in-depth analysis of the generation portfolio with the associated emission rates, vehicle mobility pattern, and employed charging mechanisms. 

In this study, we are interested in understanding how vehicle charging impacts the carbon emissions from generation plants. To mitigate carbon emissions, we propose a controlled charging strategy with an emission orientation. We evaluate both uncontrolled (direct) and controlled schemes for vehicles charging during the day or night, based on the Electric Reliability Council of Texas (ERCOT) interconnection. We assess the impacts of different charging schemes to address the following issues: 1) How much are the carbon emissions associated with EV charging? 2) What are the differences in emission when vehicles are charged during the day and night? We find that the emission results can be varied significantly driven by different charging schemes conditioning on factors of the generation mixture portfolio, the grid system load, and vehicle mobility patterns.

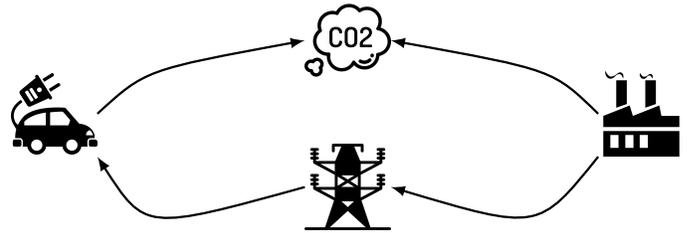
\begin{figure}[!h]
    \centering
    \begin{tikzpicture}[thick, scale=0.39, every node/.style={scale=0.39}]
    \node[inner sep=0pt] (ev) at (0, 0.5)
    {\includegraphics[width=0.16\textwidth]{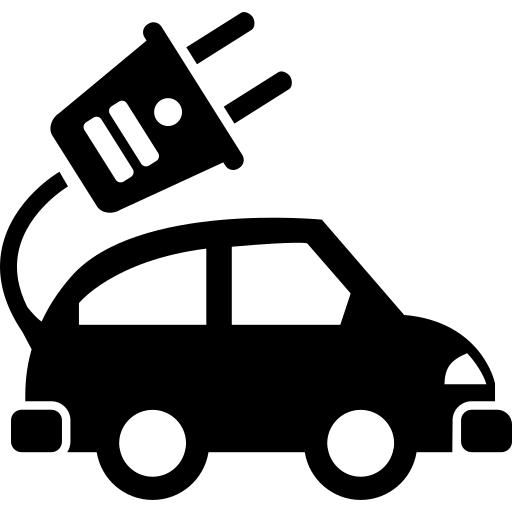}};
    \node[inner sep=0pt] (co2) at (10, 3) 
    {\includegraphics[width=0.16\textwidth]{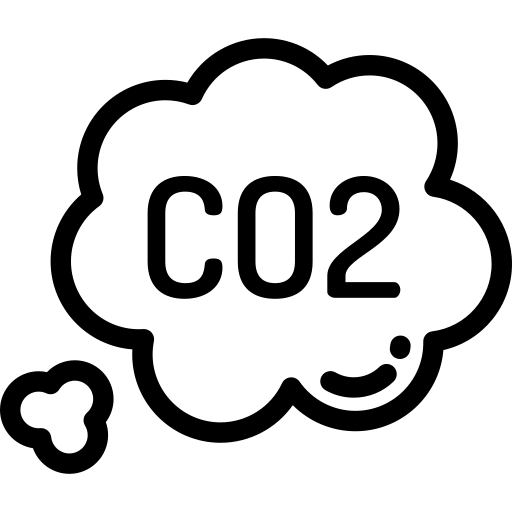}};
    \node[inner sep=0pt] (trans) at (10, -2) 
    {\includegraphics[width=0.16\textwidth]{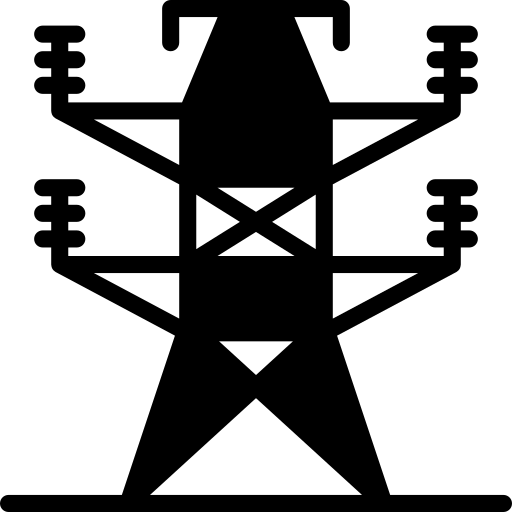}};
    \node[inner sep=0pt] (plant) at (20, 0.5)
    {\includegraphics[width=0.16\textwidth]{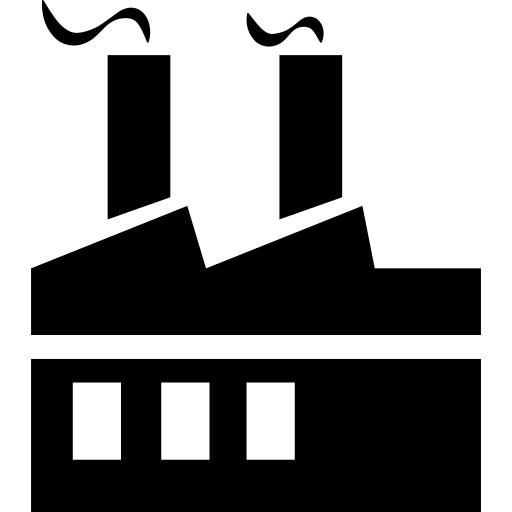}};
    \draw[->, line width=0.3mm] (ev.east) .. controls (3.1, 2) .. (co2.west);
    \draw[->, line width=0.3mm] (trans.west) .. controls (3.1, -3.5) .. (ev.south east);
    \draw[<-, line width=0.3mm] (co2.east) .. controls (17, 2) .. (plant.west);
    \draw[<-, line width=0.3mm] (trans.east) .. controls (16.5, -3.5) .. (plant.south west);
    \end{tikzpicture}
    \caption{Motivations: How does the EV charging affect the emission given the current generation setting?}
\end{figure}

\vspace{-0.12in}
\section{Related work}
Myriad studies discuss the environmental impacts of EV charging, but most neglect the fact that power plants are heterogeneous and there are different emission rates associated with these plants. Several reports that analyze the benefits of EV adoption assume that the electricity used to recharge the vehicles is generated from a particular type of power plant\cite{kliesch2006plug}. Their studies conclude that cleaner power plants yield greater environmental benefits, which is as expected. These analyses are not useful in understanding the sensitivity in emission with respect to changes in EV charging demand at specific locations on the grid and times-of-the-day. A typical study in \cite{siler2012marginal} uses a regression method to estimate marginal emissions across different
regions in the U.S. and the specific times of day. Although this is an improved method in quantifying GHG emission, it has drawbacks in that, for example, the marginal emission rate does not depend on the electricity demand estimated from regression. An another relevant work is that Denholm et al.\cite{parks2007costs} evaluate standard, delayed, off-peak, and continuous charging on emissions for Xcel in Colorado, which can be considered as an early attempt to understand the emission results due to various charging schemes. 

Similar to many analyses of marginal emission studies, our analysis contains some limitations: 1) our simulation is only suited to assess the carbon emissions given the power system structure, fuel price, vehicle technologies that was prevalent during the time when we conducted our analysis; 2) many other factors such as operating constraints, market power, or weather conditions are ignored for simplicity; 3) vehicle mobility (such as arrival/departure time) is simplified by drawing samples from a fixed mobility distribution pattern with an estimated number of charging requests. However, our main goal is to assess the carbon emissions from vehicle charging perspective. Distinguished from the previous studies, we consider EV charging demand together with system load demand and clear these total demands in the manner of economic dispatch. Our proposed charging scheme is emission friendly, and corresponding evaluations are conducted with the consideration of vehicle mobilities (day- or night-charging).     
\vspace{-0.07in}
\section{Problem formulation}
\begin{figure}[!h]
    \centering
    \begin{tikzpicture}[thick, scale=0.59, every node/.style={scale=0.59}]
    \node[inner sep=1pt] (ev1) at (0, 0.0)
    {\includegraphics[width=0.25\columnwidth]{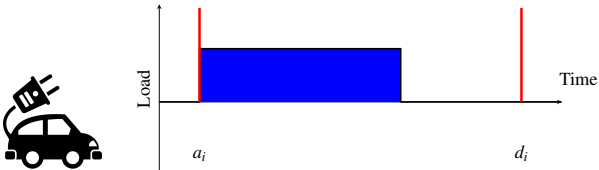}};
    \node[inner sep=35pt] (ev1node) at (ev1.south east) {}; 
    \begin{axis}[
        at={(ev1node.east)},
        width=0.5\textwidth,
        y=40mm,
        axis x line = middle,
        axis y line = left,
        ytick=\empty,
        xtick=\empty,
        yticklabels=\empty,
        xticklabels=\empty,
        ylabel={\huge Load},
        xlabel={\huge Time},
        ylabel style=
        {
          yshift=1mm, 
        },
        xlabel style=
        {
          xshift=15mm,
          yshift=5mm, 
        },
        ymin=-0.4, ymax=0.55, enlargelimits=false
        ]
        \addplot[const plot, fill=blue] coordinates {(0.0, 0)
         (0.1, 0)
         (0.1, 0.3)  
         (0.6, 0.3)
         (0.6, 0) 
         (1.0, 0)};   
        \addplot [red, ultra thick] table {
        x   y 
        0.1 0
        0.1 0.53
        };
        \addplot [red, ultra thick] table {
        x   y 
        0.9 0
        0.9 0.53
        };
        \node at (0.1, -0.35) [anchor=south] {\huge $a_i$};
        \node at (0.9, -0.35) [anchor=south]
        {\huge $d_i$}; 
    \end{axis}    
    \end{tikzpicture}
    \vspace{-11mm}
    \caption{Direct Charging}
    \label{fig:directCharging}
\end{figure}

\begin{figure}[!h]
    \centering
    \begin{tikzpicture}[thick, scale=0.59, every node/.style={scale=0.59}]
    \node[inner sep=1pt] (ev2) at (0.0, 0.0)
    {\includegraphics[width=0.25\columnwidth]{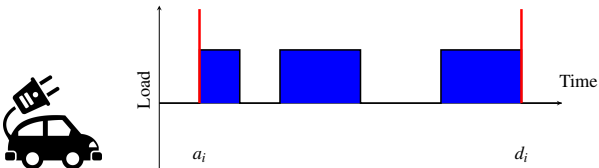}};
    \node[inner sep=35pt] (ev2node) at (ev2.south east) {}; 
    \begin{axis}[
        at={(ev2node.east)},
        width=0.5\textwidth,
        y=40mm,
        axis x line = middle,
        axis y line = left,
        ytick=\empty,
        xtick=\empty,
        yticklabels=\empty,
        xticklabels=\empty,
        ylabel={\huge Load},
        xlabel={\huge Time},
        ylabel style=
        {
          yshift=1mm, 
        },
        xlabel style=
        { 
          xshift=15mm,
          yshift=5mm, 
        },
        ymin=-0.4, ymax=0.55, enlargelimits=false
        ]
        \addplot[const plot, fill=blue] coordinates {(0.0, 0)
         (0.1, 0)
         (0.1, 0.3)  
         (0.2, 0.3)
         (0.2, 0.0)
         (0.3, 0)
         (0.3, 0.3)
         (0.5, 0)
         (0.7, 0)
         (0.7, 0.3)
         (0.9, 0.)
         (1.0, 0)
        };    
        \addplot [red, ultra thick] table {
        x   y 
        0.1 0
        0.1 0.53
        };
        \addplot [red, ultra thick] table {
        x   y 
        0.9 0
        0.9 0.53
        };
        \node at (0.1, -0.35) [anchor=south] {\huge $a_i$};
        \node at (0.9, -0.35) [anchor=south]
        {\huge $d_i$};
    \end{axis}    
    \end{tikzpicture}
    \vspace{-11mm}
    \caption{Emission-oriented Charging}
    \label{fig:EmissionCharging}
\end{figure}

Economic dispatch \cite{chowdhury1990review} is a widely accepted approach for ISOs to clear the electricity market. This approach balances the electricity generation and demand by setting the generation quantity of generators according to their marginal operation cost. We use this idea as a fundamental modeling concept throughout our analysis. Suppose an operation period contains multiple time steps indexed by $t=1, \dots, T$, where $T$ is the total number of time steps. $\mathcal{J}$ denotes a set consists of all generators indexed from $j=1, \dots, J$. Denote $q_j(t)$ as decision variables representing the generation amount (or energy output) of generator $j$ at time $t$. For simplicity we assume each generator has a time invariant  linear marginal cost $c_j$, and marginal emission rate $\theta_j$. 
We characterize each charging task (or session) 
as a 4-dimensional vector $[a, d, E, m]$ where $a$ is the arrival time, $d$ is the departure time, $E$ is the 
energy request during each charging task, and $m$ is the power capacity of a battery that 
restricts the charging speed. To match up with the time scale in the dispatch model, which has the hourly resolution, we round the arrival and departure time to the nearest hour. We may use the terms of charging tasks or charging sessions interchangeably in the rest of the paper. Moreover in the following formulation, we assume all charging sessions information is known upfront. With all the defined notations, we describe the following two charging schemes : \emph{direct charging} and \emph{emission-oriented charging}, which are depicted intuitively in Figure \ref{fig:directCharging} and Figure \ref{fig:EmissionCharging}. And the detail explanation is presented in the subsequent sections. 
%
\vspace{-4mm}

\vspace{-4mm}

\vspace{-0.05in}
\subsection{Direct Charging}
We run a direct charging scheme as the basic benchmark. The direct charging starts to deliver electricity to each vehicle right after its arrival. The charging process continues without preemption. The vehicle stops charging when the battery is full or when the vehicle leaves (whichever is earlier). The hourly charging energy of vehicle $i$ at time $t$ is $g_i(t)$. It is known according to charging tasks, as each vehicle is assumed to be charged at its maximum battery charging capability (i.e. energy delivery speed in the unit of KW). The resulting  
\begin{align*}
g_i(t) = \begin{cases}
	\min\{ \big[E_i - (t-a_i)m_i \big]_{+}, m_i \}, & \quad \text{if } a_i \leq t < d_i  \\
    0  & \quad \text{otherwise}, 
\end{cases} 
\end{align*} where $[x]_{+}=\max\{x, 0\}$. Given there are $N$ vehicles in the $T$-step horizon of operation planning, the aggregated EV demand at time $t$ can be automatically obtained, i.e. $G(t) = \sum_{i=1}^N g_i(t)$. Thus, we have the known demand $L(t) + G(t)$, where $L(t)$ is the exogenous load at time $t$. The resulting economic dispatch problem can be formulated as follows:
\begin{align}
    \min   & \sum_{t=1}^{T} \sum_{j \in \mathcal{J} } c_j q_j(t) \label{ev_em:econDisp0:obj:cost_min} \\ 
    s.t.  & \qquad    \sum_{j \in \mathcal{J}} q_j(t) = L(t) + G(t) \label{ev_em:econDisp0:constr:load_balance} \\
         & \qquad    q_j(t) \leq P_j \label{ev_em:econDisp0:constr:plant_cap} \\ 
        & \qquad -r_j \leq q_j(t+1) - q_j(t) \leq r_j  \label{ev_em:econDisp0:constr:ramping} \\      
        & \qquad q_j(t) \geq 0  \label{ev_em:econDisp0:constr:non_neg_gen},
\end{align}
where $c_j$ is the time invariant marginal cost of generator $j$, $P_j$ is the capacity of generator $j$, and $r_j$ is the ramping up/down limit of generator $j$. Notice that $c_j$, $P_j$, and $r_j$ are known parameters for the model. We have 
$r_j = 0.6 P_j$ if generator $j$ is coal fueled. Otherwise $r_j = P_j$ for other fuel-type generators.   
We set hourly ramp rate to 60\% of the capacity for coal generators\footnote{Comparing with the fuel type of Hard coal, Lignite, CCGT, and GT, we take the average ramp-up and -down speed in \cite{hentschel2016parametric}.}.  
The objective (\ref{ev_em:econDisp0:obj:cost_min}) finds the minimum marginal cost of energy for a collection of generators over the $T$-step period. Constraint (\ref{ev_em:econDisp0:constr:load_balance}) ensures the energy supply equals the demand. Constraint (\ref{ev_em:econDisp0:constr:plant_cap}) the power yielded from the generator is less than or equal to its capacity. Constraint (\ref{ev_em:econDisp0:constr:ramping}) is the ramping constraint because some type of generator cannot ramp fast enough.  The resulting emission quantity is $\sum_{t=1}^T\sum_{j \in \mathcal{J}}\theta_j q_j(t)$.

%

\subsection{Emission-oriented Charging}

To implement emission oriented charging scheme, we assume the dispatch sequence of generators follows their marginal operation cost in the ascending order. Given this clearing mechanism of the current electricity market, we minimize the total CO$_2$ emission amount over the $T$-step period. Recall that $c_j$ is the marginal operation cost of generator $j$. The the sequence $\{1, 2, ..., J\}$ is sorted according to $c_j$ in ascending order, where $J$ is the total number of dispatchable generators. In this formulation, we define $q_j(t)$, $G(t)$ and $z_j(t)$ as the decision variables representing the energy generation of generator $j$ at time $t$, the EV charging amount of energy at time $t$, and the indicator that whether generator $j$ is running at time $t$. $m_i$ is the maximum battery charging speed for vehicle $i$. $a_i$ and $d_i$ are the arrival and departure time of vehicle $i$. $E_i$ is the energy amount required by vehicle $i$. The green charge model can be expressed as     
\begin{align}
  \min  & \sum_{t=1}^{T}\sum_{j \in \mathcal{J}} \theta_j q_j(t) \\
    s.t. \qquad &   z_j(t) \in  \{0, 1 \} \\
        & q_j(t) \leq z_j(t) \min \{P_j, q_j(t-1) + r_j\} \label{ev_em:model:greenCh:constr:seq:cap1} \\
        &  q_j(t) \geq z_{j+1}(t) \min\{P_j, q_j(t-1) + r_j\} \label{ev_em:model:greenCh:constr:seq:cap2} \\
        & \sum_{j \in \mathcal{J} } q_j(t) = L(t) + G(t) \label{ev_em:model:greenCh:constr:gen_dem_balance} \\
        & G(t) \leq \sum_{i=1}^N m_i \zeta_i(t) \label{ev_em:model:greenCh:constr:bat:cap} \\
    & \zeta_{i}(t) = \begin{cases}
            1 & \quad  a_i \leq t < d_i   \\
            0 & \quad otherwise
        \end{cases}  \label{ev_em:model:greenCh:constr:arr_dep:indicate} \\
    & \sum_{t=1}^T G(t) = \sum_{i=1}^N E_i \label{ev_em:model:greenCh:constr:EV_demand_agg}   \\
    & \quad -r_j \leq q_j(t+1) - q_j(t) \leq r_j  \label{ev_em:model:greenCh:constr:rampping} \\    
    & \quad q_j(t) \geq 0  \label{ev_em:model:greenCh:constr:gen_nonneg} 
\end{align}
where $\theta_j$ is the emission rate of generator $j$.  $\zeta_i(t)$ is the parameter determined by the charging session associated with corresponding vehicle $i$. $\zeta_i(t)$ is 1 if $t$ is in the period of arrival and departure time. $r_j$ is also a parameter representing the ramping capacity for generator $j$. 
The constraints \eqref{ev_em:model:greenCh:constr:seq:cap1} and (\ref{ev_em:model:greenCh:constr:seq:cap2}) ensure the next generator $j+1$ is not utilized until the current cheaper generator $j$ is used (sometimes may be partially used because of slow ramping). The constraint (\ref{ev_em:model:greenCh:constr:gen_dem_balance}) is balancing the supply and demand of the energy. The constraint (\ref{ev_em:model:greenCh:constr:bat:cap}) guarantees that the energy provided by generators at time $t$ won't go over the aggregated charging capability at the corresponding time. The constraint (\ref{ev_em:model:greenCh:constr:EV_demand_agg}) is balancing the energy supply and demand for EVs. This is a mix-integer problem.           

The aggregated charging amount, i.e. the solution of previous problem $G^*(t)$, then needs to distribute to individual charging vehicles. We formulate this energy disperse task as another optimization that is 
\begin{align}
    & \sum_{t=1}^T \| \sum_{i=1}^N g_i(t) - G^*(t) \| \\
    s.t. & \quad \sum_{t=1}^T g_i(t) = E_i \label{ev_em:model:greenCh:disperse:c1}\\
        & 0 \leq g_i(t) \leq m_i, \quad \text{if } a_i \leq t < d_i, \label{ev_em:model:greenCh:disperse:c2} \\
				& g_i(t) = 0, \quad \text{if } t < a_i, \text{or } t \geq d_i. \label{ev_em:model:greenCh:disperse:c3} 
\end{align}
\emph{Remark:} The most straightforward implementation of the emission-oriented charging optimization is to use the constraints including $G(t) = \sum_{i=1}^N g_i(t)$, and (\ref{ev_em:model:greenCh:disperse:c1}) - (\ref{ev_em:model:greenCh:disperse:c3}) instead of   (\ref{ev_em:model:greenCh:constr:bat:cap}) and (\ref{ev_em:model:greenCh:constr:EV_demand_agg}). Yet the size $J$ usually goes to hundreds or even thousands in one independent system operator region. If we put tens of thousands of individual vehicles as the decision variables together with generator variables, the mix-integer problem is quite complex and could be slow to solve. Thus, we decouple the problem into two phases: i) solving charging amount in aggregate level; ii) determining energy amount on individual vehicles at certain times. The emission-oriented charging is distinct from the direct charging in that this charging scheme is driven by emission minimization and meanwhile captures the economic dispatch mechanism.

\subsection{Experiment data}
\begin{figure}[!thpb]
      \centering
      \framebox{\parbox{2.4in}{\includegraphics[width=0.34\textwidth]{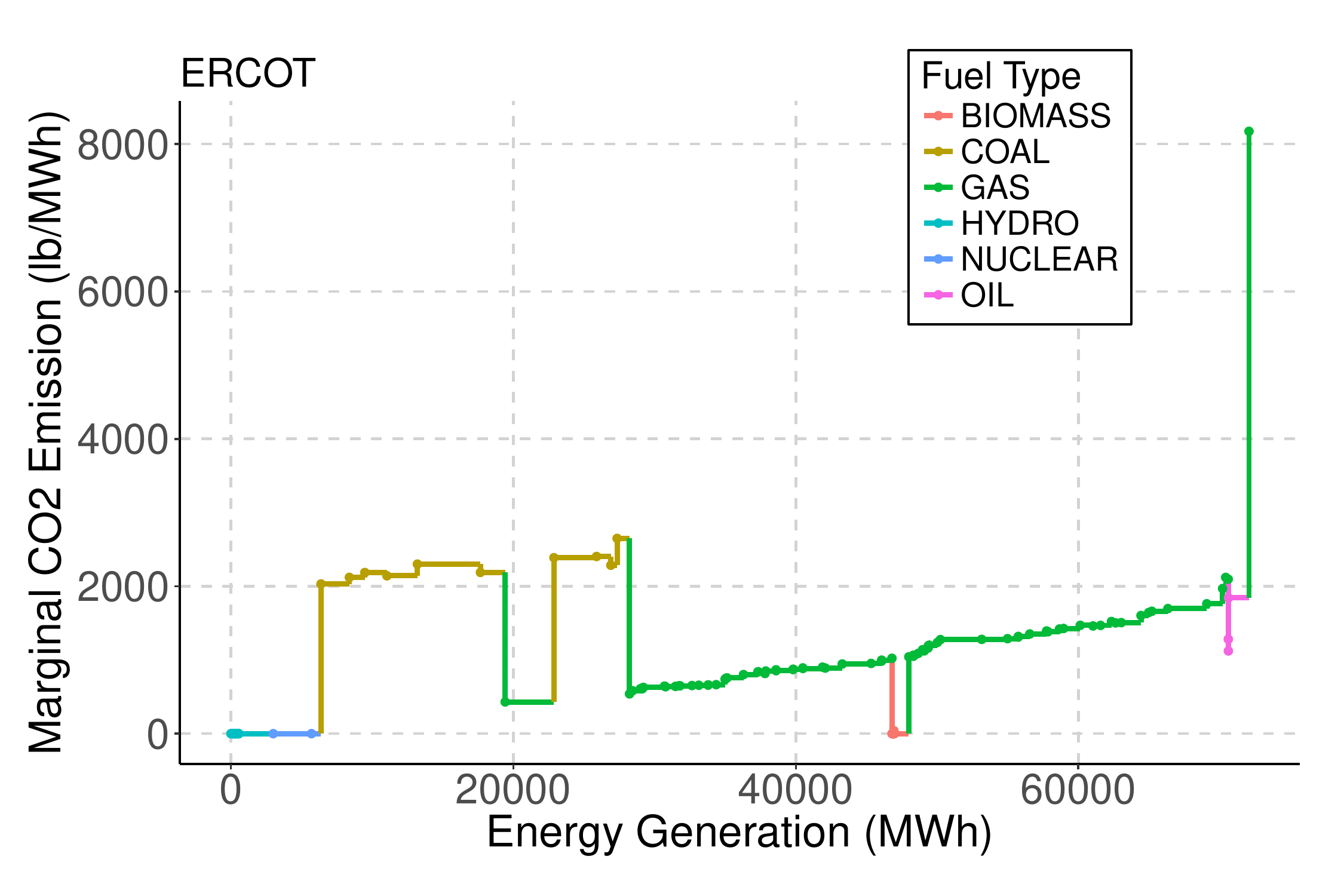}}}
      \caption{Marginal emission rate curve in ERCOT when power plants are sorted by the fuel cost.  Below 20 GW threshold, the generation is mainly coal powered.}
      \label{EV_emission:fig:co2emission:marginal_rate:curve}
\end{figure}

\begin{figure}[!thpb]
      \centering
      \framebox{\parbox{2.3in}{\includegraphics[width=0.33\textwidth]{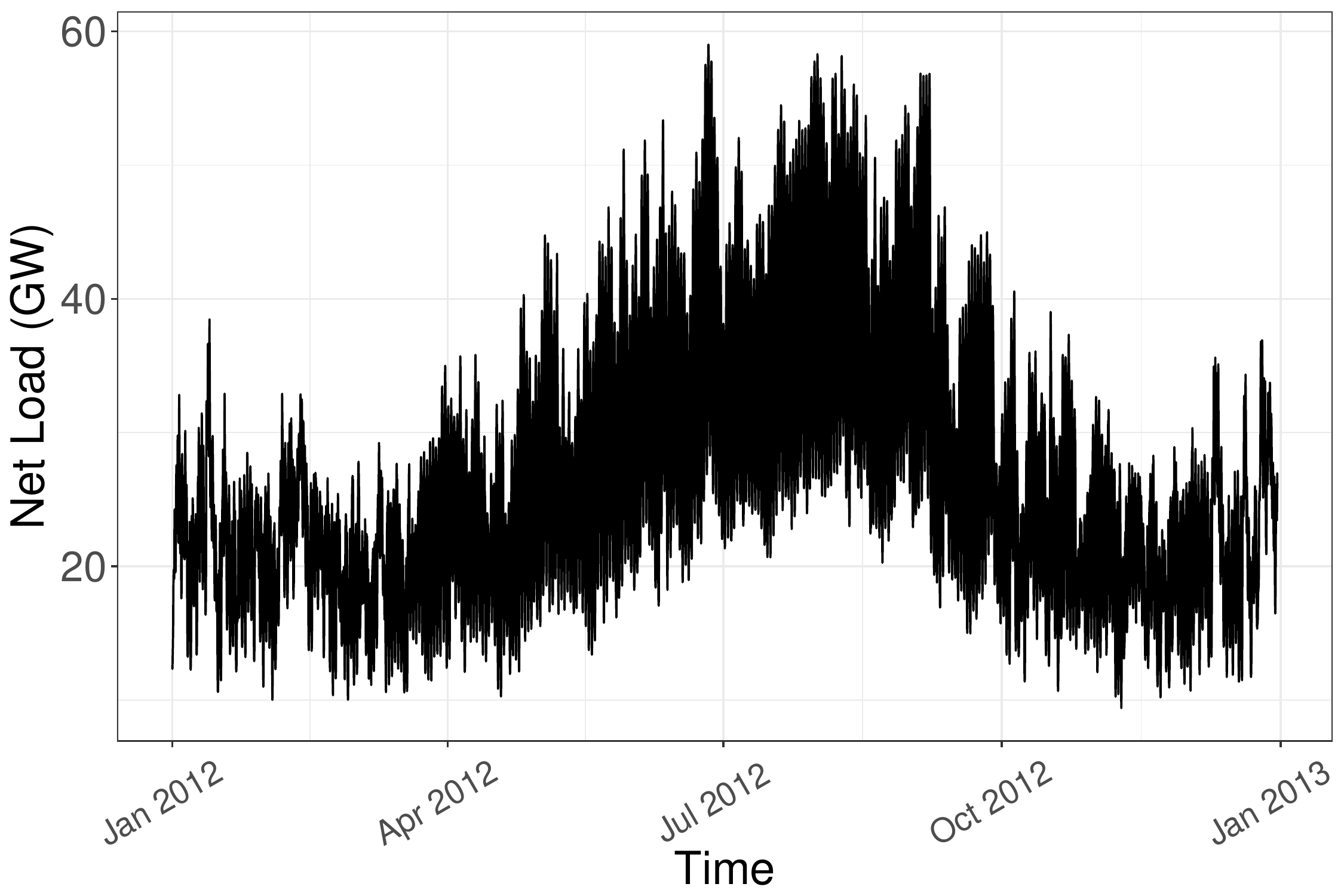} } }
      \caption{Time series of hourly net load amount in ERCOT in year 2012.}
      \label{EV_emission:fig:netload:ercot:ts:2012}
\end{figure}

\begin{figure}[!thpb]
      \centering
      \framebox{\parbox{2.3in}{\includegraphics[width=0.33\textwidth]{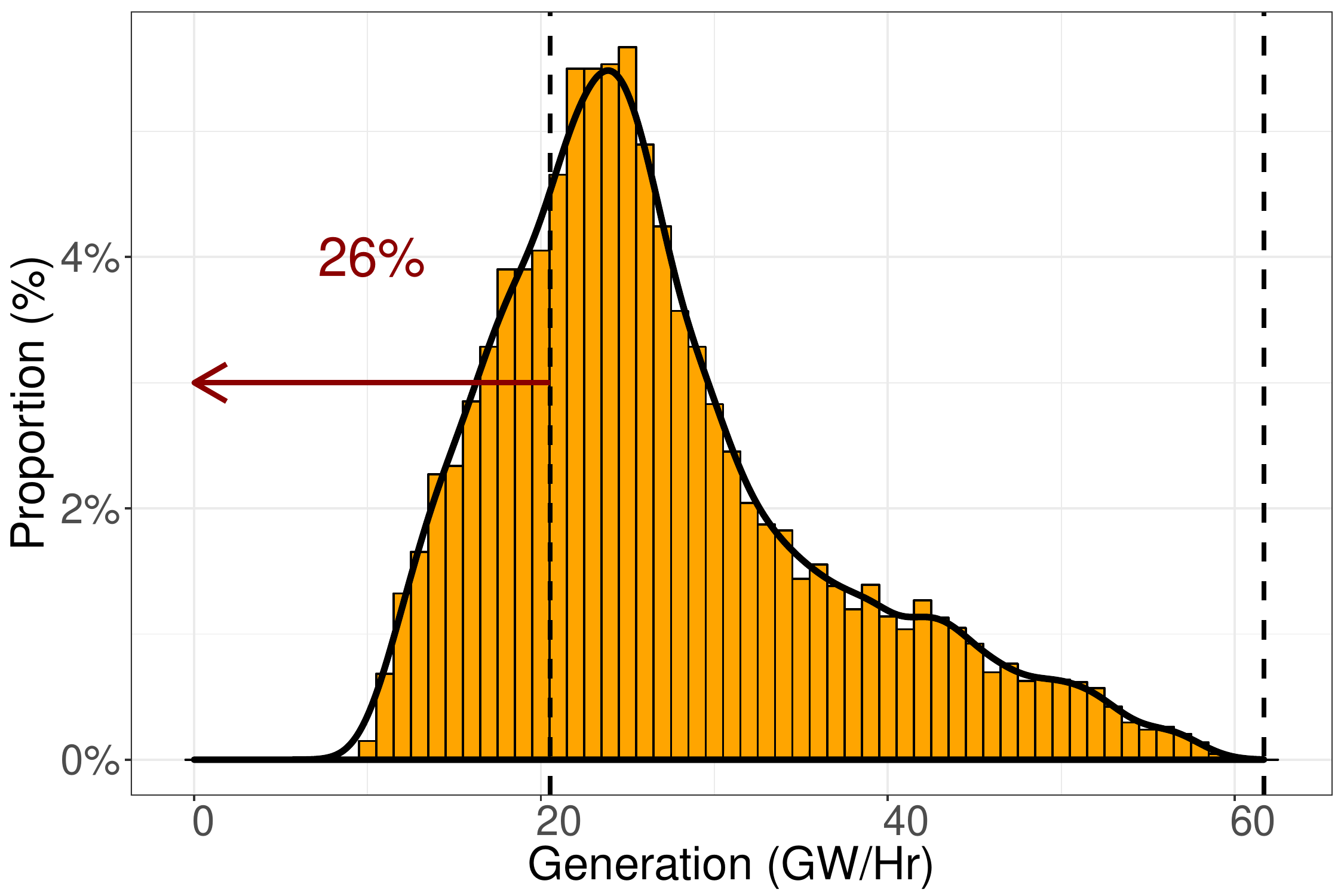} } }
      \caption{Histogram of the hourly generation amount in ERCOT over a year without considering EV charging; Two dashed vertical lines are derived from piecewise linear fitting break points; There are 26\% hours out of 8760 hours in a year that has generation amount lower than 20 GW  }\vspace{-0.14in}
      \label{EV_emission:fig:netload:ercot:histogram:2012}
\end{figure}

We assume the marginal operation cost can be approximated by fuel cost, as it plays an important role in daily operations. The associated energy source prices, including fuel, coal, and gas are quoted from Environmental Protection Agency (EPA)\footnote{We use the price data in year 2012 from {https://www.eia.gov/}}. The Emission rates of CO$_2$ are obtained from Emissions \& Generation Resource Integrated Database (eGRID). Fig~\ref{EV_emission:fig:co2emission:marginal_rate:curve} depicts the marginal emission curve. 
The load generation data is from Electric Reliability Council of Texas (ERCOT) in year 2012. Fig \ref{EV_emission:fig:netload:ercot:ts:2012} shows the net load time series over the course of the year. Fig \ref{EV_emission:fig:netload:ercot:histogram:2012} shows the hourly generation histogram over the year 2012. There are 26\% of the hours with generation amount lower than 20 GW\footnote{We offset the wind and solar generation by subtracting the existing load, as the marginal cost of renewable generation is almost negligible. }.
%
%

\begin{figure}[!thpb]
      \centering
      \framebox{\parbox{2.5in}{\includegraphics[width=0.35\textwidth]{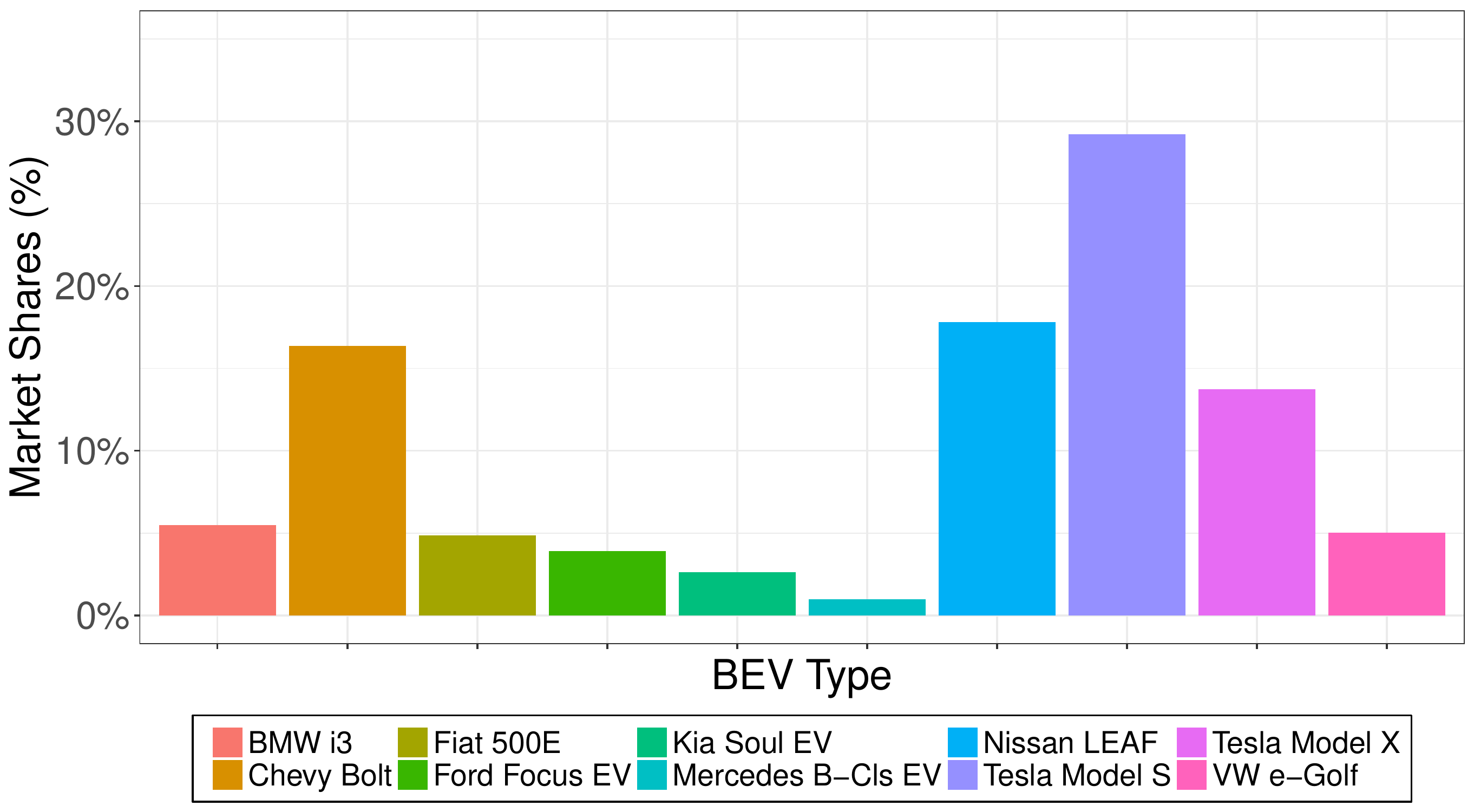} } }
      \caption{Battery Electric Vehicles Market Shares in US \cite{zhou2016light}}\vspace{-0.14in}
      \label{EV_emission:fig:EV_marketshare:2017}
\end{figure}

The electric vehicle market share and sales volume data are obtained from a study at the Argonne National Laboratory~\cite{zhou2016light}. We consider the Battery Electric Vehicle for simplicity, which accounts for 48\% of total plug-in vehicles (PEV) sales. Until early 2017, the total sales volume of PEV in accumulation is around 580,000 in US. To roughly estimate the EV number in Texas, we use the ratio of Texas Gross domestic product (GDP), i.e. $1.648$ trillion, versus the U.S GDP, i.e. $17.914$ trillion, as a scale factor. We take GDP ratio as a rough approximation factor due to the high positive correlation between GDP and vehicles volumes \cite{davis20162015}. Correspondingly, we set the 25000 EVs in our simulation ( $25000 \approx 580000 \times 0.48 \times (1.648/17.914) $ ). The vehicle travel pattern is simulated according to the typical travel behavior of home-to-work commuters  \cite{Xu2018PlanningEVcharging, chen2016parking}. These studies show that typical arrival times at workplaces ranges from 7am-10am. Departure times from workplaces are mainly from 4pm-8pm. Hence we simulate all samples of EV charging sessions from a multinomial distribution which is derived from the market shares depicted in Fig.~\ref{EV_emission:fig:EV_marketshare:2017}, and set the arrival and departure time uniformly between 7am-10am (4pm-8pm for night-charging) and 4pm-8pm (7am-10am for night-charging) respectively. The charging requirement is the maximum amount of energy that can be delivered to the vehicle (i.e. the corresponding battery capacity or the parking duration times the max energy delivery rate). The charging delivery rate is accounted as Level II charge as the maximum rate, which needs 240-volt chargers. Table \ref{EV_emission:EV:battery:standards} describes the standards of the major BEVs battery capacity and charging speed. For the simplicity of simulation comparisons, we sample the same amount of total charging demand during the day or at night.                 
%
\vspace*{-0.04in}
\vspace*{-0.05in}
\begin{table}[!hpbt]
\caption{Typical PEV battery capacity and charging speed}
\label{EV_emission:EV:battery:standards}
\centering
\begin{tabular}{|c||c|c|}
\hline
Vehicle & {Battery} & Max Delivery   \\
& Capacity (kWh) & Rate (kW) \\
\hline
BMW i3 (2015) & 22 & 7.7 \\
Chevrolet Bolt  &  60 & 7.7  \\
Fiat 500E & 24 & 6.6 \\
Ford Focus EV (2012) & 23  & 6.6 \\
Kia Soul EV & 27 & 7.7 \\
Mercedes B-Cls EV &  28 & 11 \\ 
Nissan Leaf (2016) & 26  & 6.6  \\
Tesla Model S &  85  & 11 \\
Tesla Model X & 85 & 11\\
Volkswagen e-Golf & 30 & 7.7  \\
\hline 
\end{tabular}
\vspace{-0.14in}
\end{table}
%
%
%
%
\vspace{-0.05in}
\section{Results}
\begin{figure}[!thpb]
      \centering
      \framebox{\parbox{2.5in}{\includegraphics[width=0.35\textwidth]{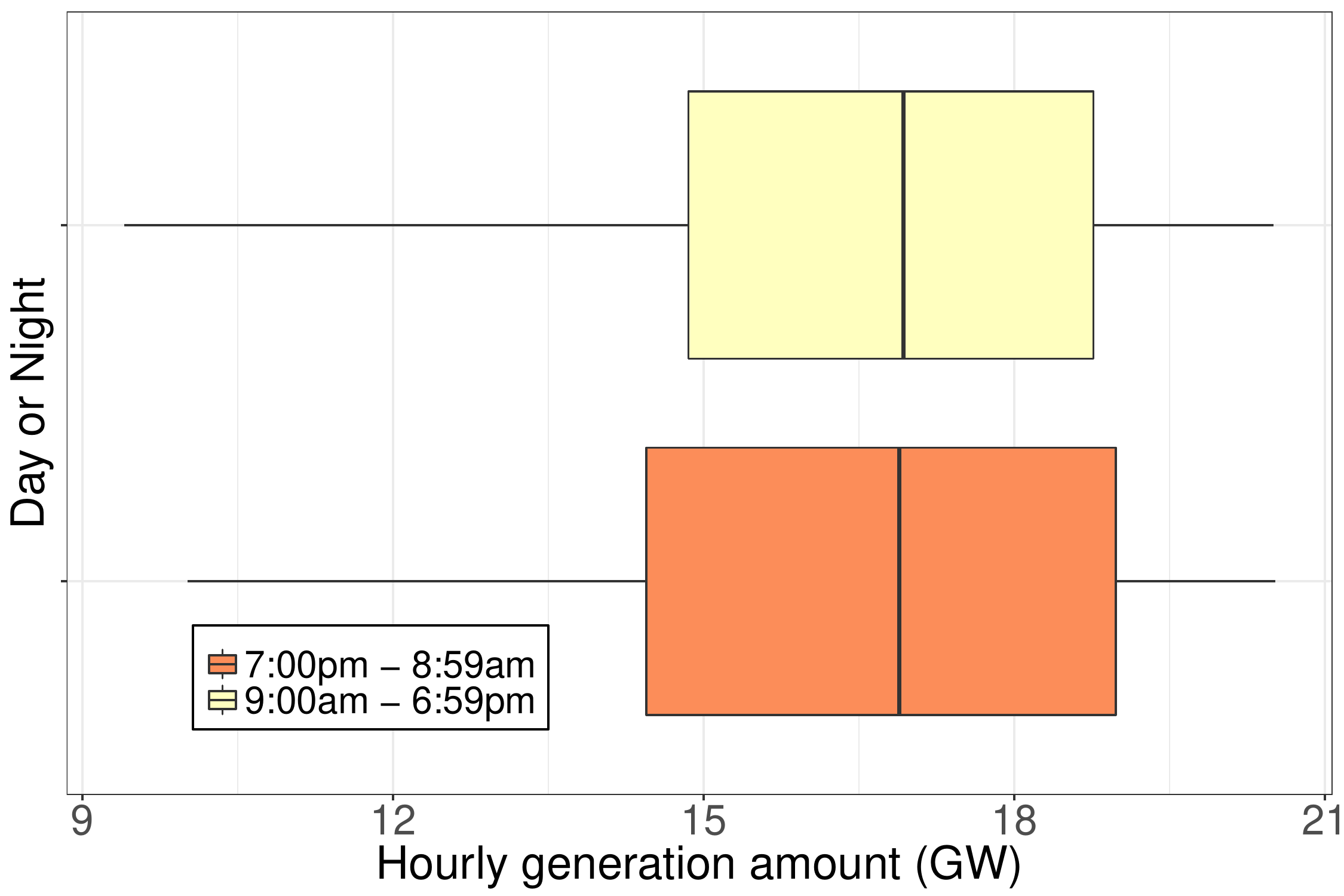} } }
      \caption{The comparison of hourly generation quantity in the day and night. The mean of day time hourly generation is around 16.70 GW, which is larger than the mean of night time hourly generation that is 16.57 GW  } \vspace{-0.17in} \label{EV_emission:fig:lowHourlyGen:boxplot:DnN:ercot}
\end{figure}

We first discover that 20 GW is one critical point from Fig~\ref{EV_emission:fig:co2emission:marginal_rate:curve}, as the marginal power resource is mainly coal-fueled when the energy generation is below that threshold. Given the low generation threshold 20 GW, we further take a look at the distribution of low generation hours. Apart from the weekends, people usually think the low generation hours will mainly be at night because the evening load consumption is usually smaller than the day load consumption. Fig \ref{EV_emission:fig:lowHourlyGen:boxplot:DnN:ercot} indicates that the nighttime load has a slightly lower total energy than the daytime load. However, we find that the number of low generation hours in the daytime (around 1000) is more than those at night (less than 750), even though the day time span is shorter than the night time span in a day. Such a pattern allows us to have the hypothesis that by allocating the charging demand intelligently in the day will yield significant difference on marginal emissions comparing with direct charging. This is also supported by our later investigation that reveals day charging is more emission friendly than night charging.  

%

%
%
%
%
\begin{figure}[!thpb]
      \centering
      \framebox{\parbox{2.5in}{\includegraphics[width=0.35\textwidth]{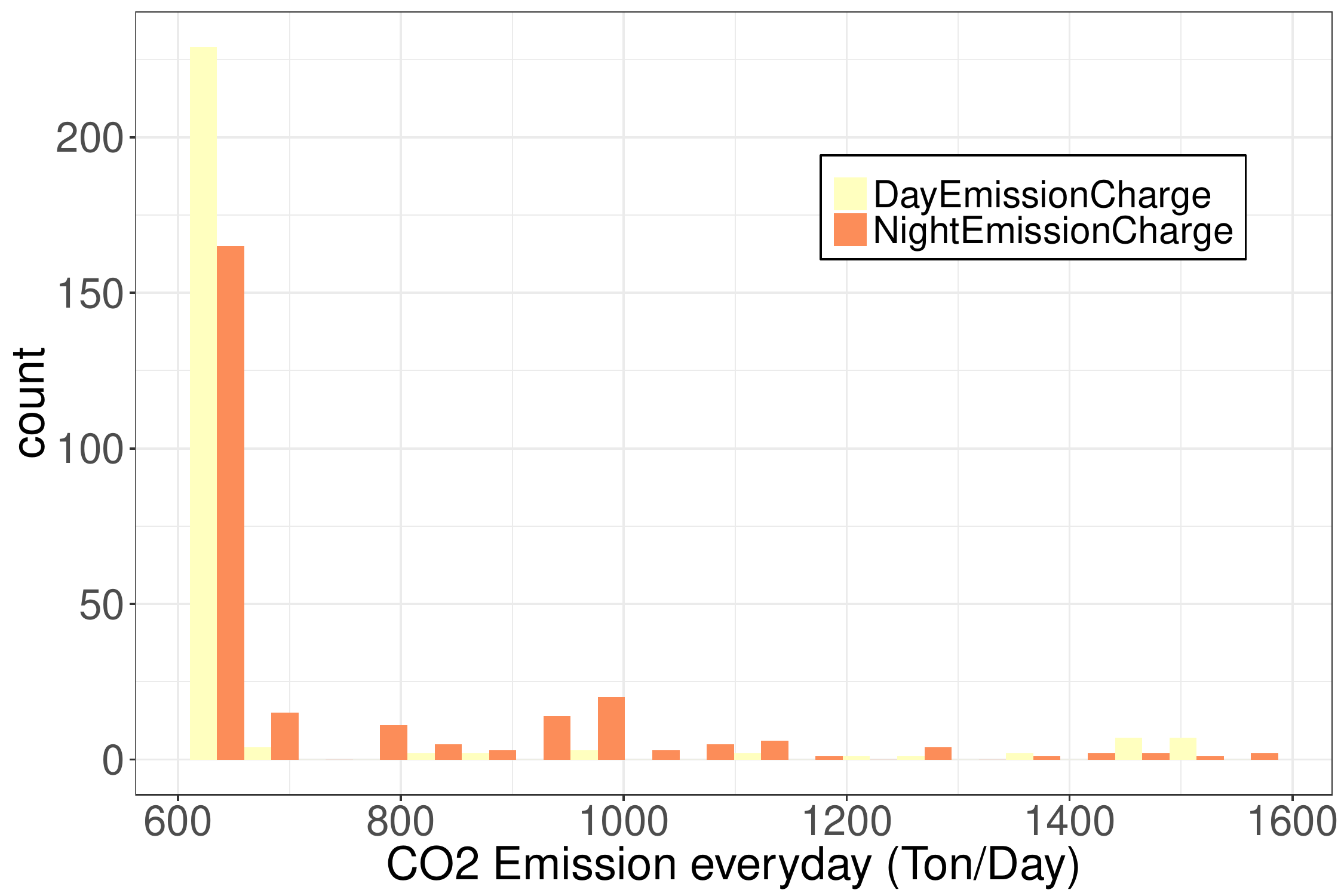}}}
      \caption{Given both approaches are emission oriented, the day-charging has more days that have low emissions than night-charging in a year}\vspace{-0.1in}
  \label{EV_emission:fig:emQ_ByCharging:hist:Count:DnN:ercot}
\end{figure}

\begin{figure}[thpb]
      \centering
      \framebox{\parbox{2.4in}{\includegraphics[width=0.33\textwidth]{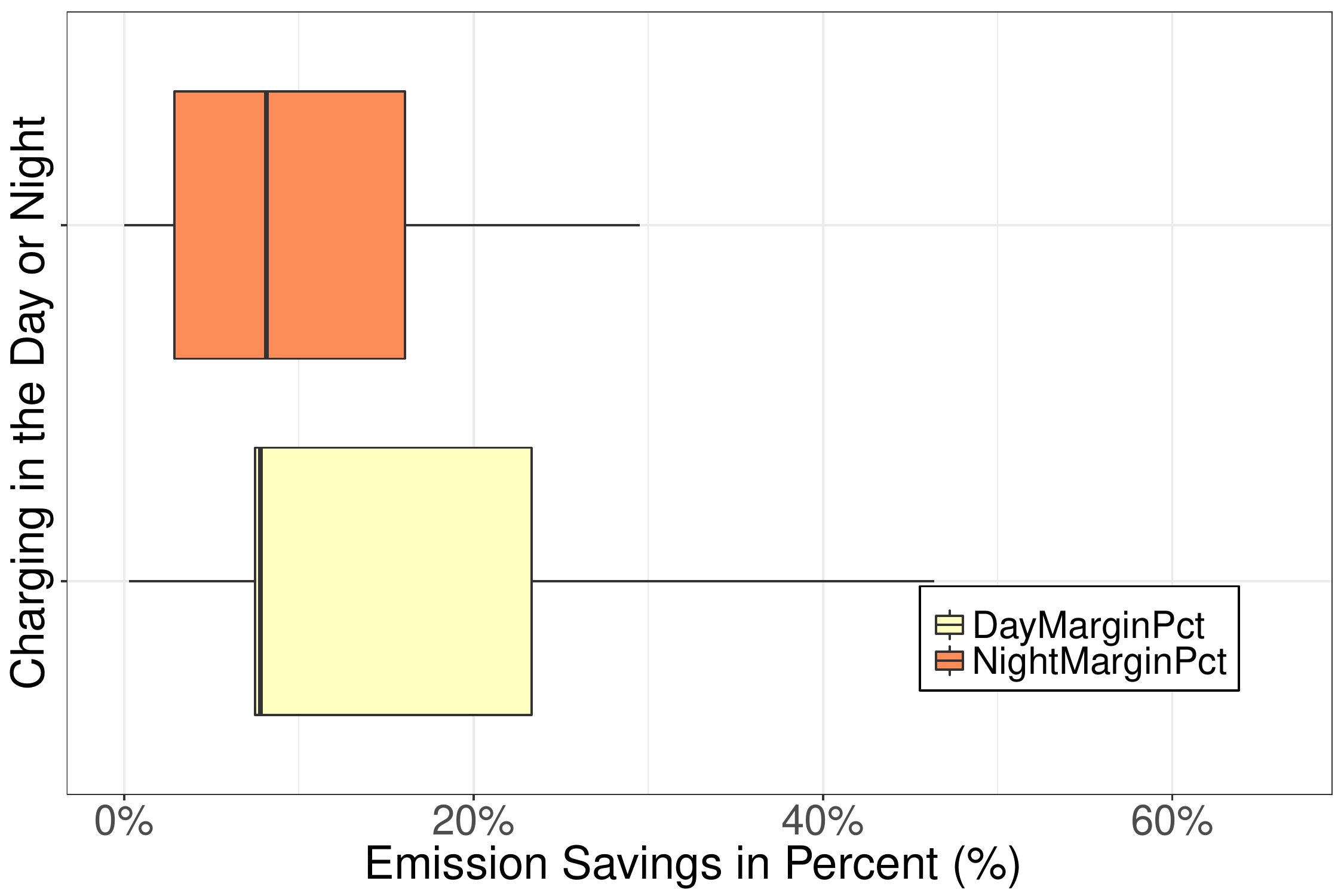}}}
      \caption{We show the difference between direct charging and emission oriented charging during the daytime versus nights; The x-axis is the ( (CO$_2^{emission charging}$ - CO$_2^{direct charging}$ )/CO$_2^{direct charging}$); Within a year, 30\% days have a significant difference of CO$_2$ emissions when applying emission oriented charging vs direct charging. When applying emission oriented charging during the night, we reach about 12.1\% CO$_2$ emission savings compared with the direct charging. Whereas we can obtain about 13.8\% CO$_2$ emission savings on average, and in many cases even more, when applying emission charging compared to the direct charging during the daytime }\vspace{-0.07in}
      \label{EV_emission:fig:EmChargeVsDirectCharge}
\end{figure}

\begin{figure}[!thpb]
      \centering
      \framebox{\parbox{2.4in}{\includegraphics[width=0.33\textwidth]{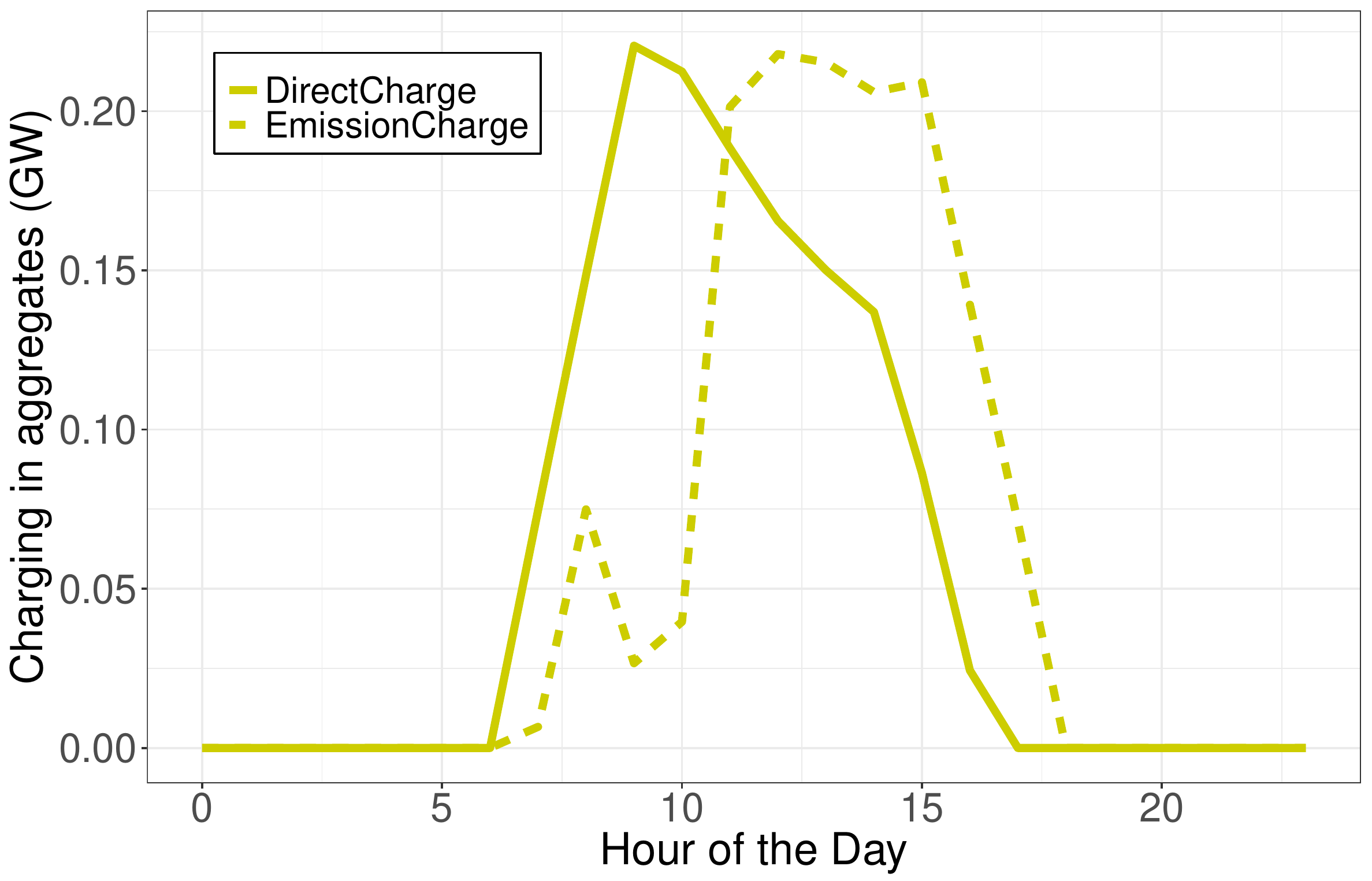}}}
      \caption{EVs charging amount in aggregates during the day}
      \label{EV_emission:fig:chargePattern:aggregates:Day}
\end{figure}

\begin{figure}[!thpb]
      \centering
      \framebox{\parbox{2.4in}{\includegraphics[width=0.34\textwidth]{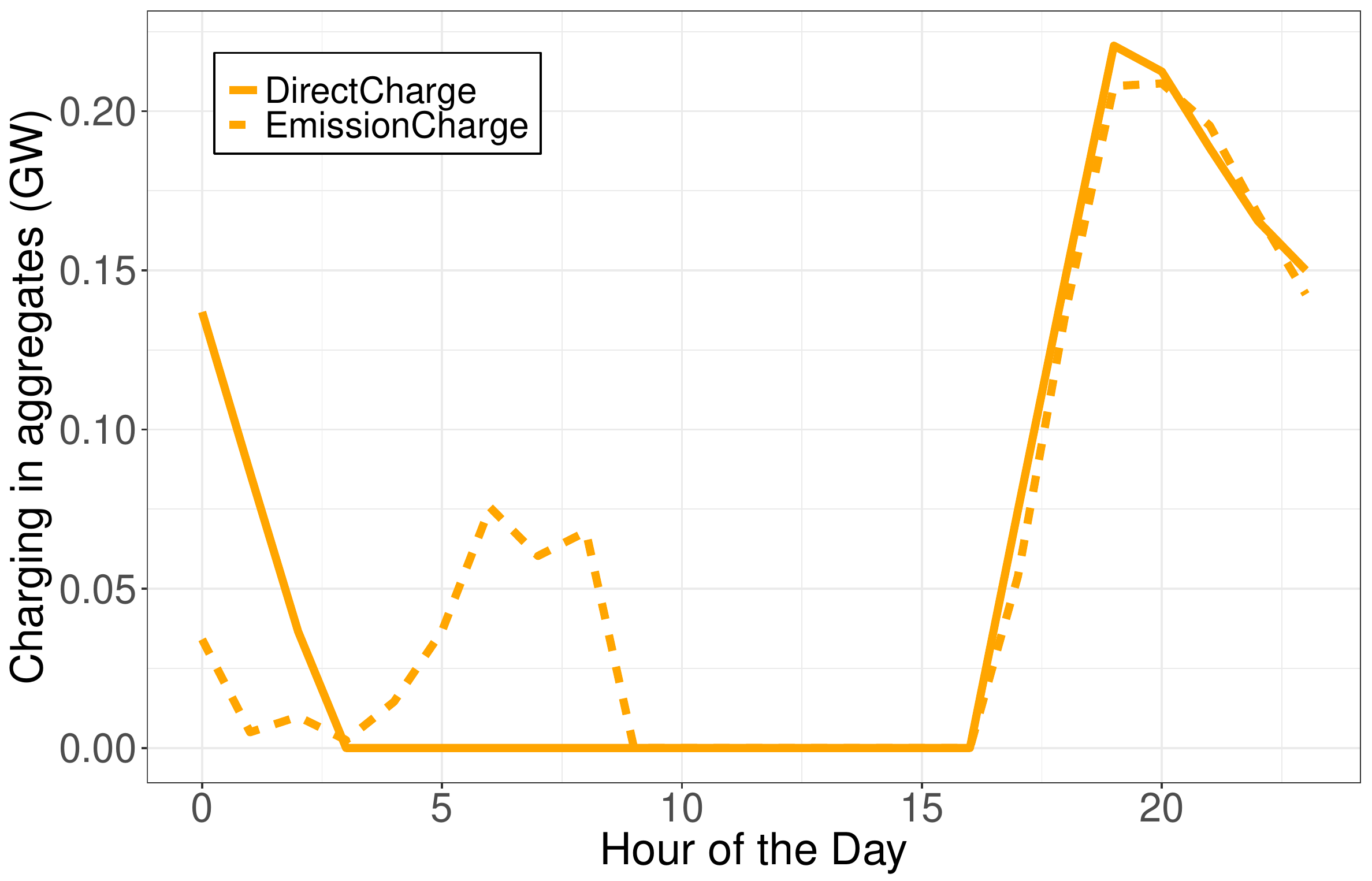}}}
      \caption{EVs charging amount in aggregates during the night}\vspace{-0.15in}
      \label{EV_emission:fig:chargePattern:aggregates:Night}
\end{figure}

%
 We compare the results of \emph{emission-oriented charging} in the day and night on weekdays, because the vehicle mobility pattens on weekends are significantly different from the arrival and departure time distribution that we previously assumed. The daily CO$_2$ emission quantity over a year is given in Fig \ref{EV_emission:fig:emQ_ByCharging:hist:Count:DnN:ercot}. During the weekdays in 2012, the emission amount largely concentrates around 626 US-ton per day for the day-time emission-oriented charging. On the other hand the night-time emission-oriented charging yields more emissions scattered from 626 ton per day up to nearly 1600 ton per day. The \emph{emission-oriented charging} at night has some days that produce a significantly high amount of emissions due to the fact that the hourly base load\footnote{base load refers to the load without EV charging demand. } at night is lower than the hourly base load during the day. Such a fact indicates that the grid system would dispatch low marginal but high emission cost generation resource first to compensate the charging demand at night. For example coal generators will usually be a good candidate to support night-time charging demand when the system generation level is relatively low.              
We also take a further step to compare the emission savings of applying the \emph{emission-oriented charging} versus the direct charging. Within a year, 30\% of days have a significant difference\footnote{the difference of emissions is larger than 0.01 ton per day} when applying two different charging schemes. Fig.~\ref{EV_emission:fig:EmChargeVsDirectCharge} shows that the emission-oriented charging in the day can yield 13.8\% CO$_2$ emission savings compared to the direct charging in the day. Whereas the \emph{emission-oriented charging} during the night has 12.1\% CO$_2$ emission savings compared to the corresponding direct charging at night. Meanwhile the CO$_2$ emission savings by charging in the day has a heavier upper tail which can reach to 20\% savings or more in some days (Fig.~\ref{EV_emission:fig:EmChargeVsDirectCharge}). But the CO$_2$ emission savings of charging at night ranges only from 5\% to 15\% for most days. Overall, we find that emission-oriented charging reduces significantly more CO$_2$ emissions compared with direct charging either during the day or night. Applying emission-oriented charging has lower emissions than using this charging method during the day than at night. Emission-oriented charging during the daytime yields higher CO$_2$ emission savings than using this method at night. 
%
The corresponding daily charging pattern in aggregates for both emission oriented charging and direct charging during the day and night time is displayed in Fig \ref{EV_emission:fig:chargePattern:aggregates:Day} and Fig \ref{EV_emission:fig:chargePattern:aggregates:Night} respectively. These are the averaged curves across different weekdays days over a year. We notice, during the day, the charging profile of emission-oriented charging tends to defer the energy requests to some later hours. We conjecture that such a pattern is closely related to the marginal generation of the system. Because high system load often appears after 10am which require more gas plants to contribute the power generations. Similar reason can be inferred from the small hump in Fig.~\ref{EV_emission:fig:chargePattern:aggregates:Night} of the emission oriented charging profile. 
%
%
%
%


%
%
%
%
%
\vspace*{-0.1in}
\section{Conclusions and Discussions}
According to the current energy market, a significant share of emission-related benefits remain external to electricity transactions. Policies that are designed to support socially efficient level of deployment of new technologies (such as EVs) should consider incentives that accurately reflect external uncompensated benefits. In this paper, we evaluate carbon-emission savings generated by a emission-oriented charging scheme during the day or night. Our simulation results suggest that certain policy incentives regarding EV charging should be taken into account, because CO$_2$ emissions differ significantly with respect to charging period in a region. In our future research, we will assess how different charging scenarios can affect the total system load at various EV penetration levels with realistic vehicle mobility patterns and incorporation of renewable resources such as solar and wind.

\addtolength{\textheight}{-12cm}   






%
%


%
%
%
%

%
%

\bibliography{EV_Charge_Emission}	
\bibliographystyle{ieeepes}

\end{document}